\documentstyle[11pt,moriond]{article}

\def\be{\begin{equation}}
\def\ee{\end{equation}}
\def\bea{\begin{eqnarray}}
\def\eea{\end{eqnarray}}

\begin{document}
\vspace*{4cm}
\title{HIGH ENERGY PLASMA PHYSICS \\
IN BLACK HOLE ENVIRONMENT WITH JETS}

\author{ Guy PELLETIER and Gilles HENRI }

\address{Laboratoire d'Astrophysique, Observatoire de Grenoble,\\
B.P. 53, 38041 Grenoble Cedex 09, France}

\maketitle\abstracts{
The Compton Gamma Ray Observatory revealed the high energy emission of
some Active Galactic Nuclei called "Blazars"; few of them have also been
observed by ground Cererenkov arrays. Those sources observed with Whipple
Observatory emit gamma rays up to few TeV. From the spectral data and
variability measurements, a purely electrodynamical description of this
emission can be proposed, provided that it comes from relativistic flares
at the beginning of the jet. Because of the source compacity, the creation
of electron-positron pairs and their escape from the black hole environment
could likely explain the shape of the spectra. A plasma dominated by
relativistic electrons (positrons) close to equipartition with the
magnetic field is a suitable acceleration medium that can account for
both the required high Lorentz factors and the short variability.}

\section{Pair creation}

Active Galactic Nuclei are compact sources in a precise physical sense,
for they display an intense X-radiation field that makes them optically
thick to gamma-rays, at least within some radius of order $100 r_G$ ($r_G
= 2GM_*/c^2$ is the gravitational radius of the presumed black hole).
This is measured by a number called "compactness" and
defined for a size $R$ by:
\begin{equation}
        l \equiv \frac{\sigma_T L_X}{4\pi R m_e c^3} \ .
        \label{CM}
\end{equation}
So when $l\gg 1$, every gamma photon interacts with an X-photon to give a
pair of electron-positron. There are several possible processes that
produce gamma photons and thus pairs in AGNs. They can be a consequence of
the Penrose mechanism, the effect of a gap parallel electric field in the
vicinity of a rotating black hole, of the Fermi acceleration mechanisms
that maintain hadronic collisions (proton-proton collisions and
proton-soft photon collisions) and the inverse Compton process. The
observed gamma radiation cannot escape from the central region. However
the gamma emission is observed only in some blazars (Von Montigny et al.
1995) that have jets with
superluminal motions and is understood if one takes into account the
Doppler beaming. Indeed the relativistic motion (of bulk Lorentz factor
$\gamma_b$) of the emitting cloud
along a direction having
a rather small angle $\theta$ with respect to the line of sight produces
relativistic aberrations that depends on the Doppler factor ${\cal D}$:
\begin{equation}
        {\cal D} \equiv \frac{1}{\gamma_b(1-\beta_b cos \theta)}
        \label{DF}
\end{equation}
Superluminal motions indicate that the Doppler factor is large and thus
the radiation is beamed towards the observer according to:
\begin{equation}
        I(\omega, \theta) = {\cal D}^3 I_0(\frac{\omega}{{\cal D}}) \ ,
        \label{AL}
\end{equation}
and thus the luminosity appears much larger than its intrinsic value which, in
fact, remains smaller than the UV bump. Moreover the variability time
scale appears shorter by a Doppler factor. Since the variability time
gives the maximum size of the source, the gamma emission region is not
larger than $100 r_G$.

So the natural understanding of the high energy emission of blazars is
to consider that it comes from the region where relativistic clouds become
optically thin to gamma rays at about $100 r_G$ (Henri, Pelletier,
Roland 1993).  A detailed calculation of the radiative transfer of the
high energy photons (X and gamma) emitted by the electrons of a
relativistic jet in the anisotropic radiation field of
an accretion disk has been performed (Marcowith, Henri, Pelletier 1995).
It takes into account the Inverse Compton process on the UV bump and the
pair creation process by two gamma photons, the X- and $\gamma$-rays
being generated by the Inverse Compton process and there is also a small
contribution of the annihilation. Because of the stratification and the
growth of the pair density up to a value that makes the source optically
thick to Thomson scattering of the soft photons, the
spectrum breaks around few MeV, the higher $\gamma$-rays being still in the
optically thick regime, whereas the X-rays are in optically thin regime
to $\gamma \gamma$-pair production. So far this model is the only one
that accounts for the
observed spectrum break, since it predicts that the gamma index is twice
the X index, whereas the incomplete Compton cooling model (Dermer and
Schlikheiser 1992) predicts the canonical $1/2$ steepening. A pair model
has been proposed also by Blanford and Levinson (Blandford and Levinson
1995), but they considered the pair creation with the UV-photons only
(whereas our model takes mostly into account the creation by two gamma
photons) and they did not consider the reacceleration of the pairs, which
is an important aspect of our model. Indeed reacceleration of pairs tends
to produce a pair creation catastrophe that could likely explain the fast
variability.

\section{The acceleration processes}

The acceleration of particles in astrophysical media depends on large
magnetic disturbancies. As long as the electric field associated with
these disturbancies is neglected, the suprathermal particles suffer
pitch angle
variations only since the magnetic field "does not work". Thus a plasma
having random magnetic disturbancies plays the role of a scattering
medium for the particles characterized by a pitch angle scattering
frequency $\nu_s$ (generally much smaller than the synchrotron frequency
at the same energy $\omega_s$).

The most easily excited magnetic perturbations are in the form of Alfven
waves, that are electromagnetic waves that propagate along the average
magnetic field with a phase velocity $V_A$. The electric component is
smaller than the magnetic one by a factor $V_A/c$ so that the electric
force acting on a particle of velocity $v$ is smaller than the magnetic
force by a factor $V_A/v$. The relative rate of change of the energy is
then smaller than the relative rate of change of the pitch angle by a
factor $V_A^2/v^2$. So a source of so-called second order Fermi
acceleration is a region of space where a plasma has electromagnetic
random disturbancies that maintain a diffusion in energy space, and
the change in momentum $\Delta p$ during a time $\Delta t$ larger than
the correlation time is such that
\begin{equation}
        \frac{<\Delta p^2>}{2\Delta t} \sim \frac{V_A^2}{v^2} \nu_s p^2 \ .
        \label{DE}
\end{equation}
The second order Fermi process has an acceleration rate
$\nu_2 \sim \frac{V_A^2}{v^2} \nu_s$.

It is often said that the first order Fermi process at shocks is more
efficient; this is not plainly true. When a shock develops in a
scattering medium, the particles undergo elastic "collisions" with
upstream disturbancies that flow at a supersonic velocity $u_1$, whereas
they collide with disturbancies that flow at a subsonic velocity $u_2$
downstream. They gain energy as they cross the shock front either from
upstream or from downstream. The average (over pitch angles) momentum gain
at each cycle is
\begin{equation}
        \delta p = \frac{4}{3} \frac{u_1-u_2}{v}p
        \label{GE}
\end{equation}
The acceleration rate depends on the frequency of shock crossing.
A particle
having a velocity $v$ upstream that crosses the shock front has a
probability $\eta = 4u_2/v \ll 1$ to escape downstream (Bell 1978). Thus
the crossing frequency is $1/\eta t_e$ where $t_e$ is the average time
that a particle stays in the diffusion region in the vicinity of the
shock before escaping. That time is $t_e \sim (v^2/u_2^2) \nu_s^{-1}$;
which implies a crossing rate smaller than the diffusion rate
$\nu_c \sim (u_2/v) \nu_s$.  The acceleration rate at shock is thus given by
\begin{equation}
        \frac{<\Delta p>}{\Delta t} \sim \frac{(u_1-u_2)u_2}{v^2} \nu_s p
        \label{GA}
\end{equation}
This acceleration rate is not significantly larger than the second order
Fermi one (Jones 1994). Its major interest is that the ratio of the
acceleration rate $\nu_1$
over the escape rate is a number that depends on the shock compression
ratio only: $\nu_1 t_e = (r-1)/3$ with $r = u_1/u_2$ close to $4$ in a
strong non relativistic adiabatic shock. Which leads to the formation of
a power law energy spectrum behind the shock that is proportional to
$\varepsilon^{-2}$.

Regarding the generation of high energy gamma rays at the beginning of the
jets, or the neutrino emission in the nucleus, it is not at all useful to
invoke large shocks!.. Large shocks are not expected in these regions.
So we can invoke simply the second order Fermi acceleration in the
central region and keep the large shocks for the jet hot spots.

What are the best conditions to get a fast acceleration? Clearly the
fastest acceleration process takes place in a relativistic plasma that
has an Alfven velocity close to the velocity of light. The modified
Alfven velocity in a relativistic plasma is given by:
\begin{equation}
        V_* = \frac{C}{\sqrt{1+2\frac{P}{P_m}}} \ ,
        \label{VM}
\end{equation}
where $P$ is the relativistic pressure and $P_m$ is the magnetic pressure.
At equipartition the modified Alfven velocity equals the relativistic
sound velocity $C/\sqrt{3}$. Since these plasmas are supposed to be
magnetically confined, the propagation velocity of the electromagnetic waves
is close to the velocity of light. Under those conditions, pitch angle
scattering and acceleration work with the same time scale and the usual
expansion in power of $V_A/v$ cannot be done. Moreover
the second order Fermi process is efficient and one does not
know whether the first order Fermi process works at shocks (Baring, these
proceedings).

\subsection{Electron relativistic plasmas}

In the Fermi processes there is an essential microphysics ingredient,
namely the pitch angle scattering frequency. For high energy particles, the
only
efficient scattering process comes from the resonant interaction of these
particles with Alfven waves, which occurs with the waves having a
wavelength almost equal to the Larmor radius of the scattered particle.
In ordinary plasmas, the most massive component is due to non relativistic
protons and the
Alfven waves develop at wavelengthes larger than $V_A/\omega_{cp}$
($\omega_{cp}$ is the cyclotron pulsation of the non relativistic protons).
This puts a severe threshold for resonant interaction, especially for the
electrons that must be very energetic already: $p > m_pV_A$. However all
the single charged relativistic particles having the same momentum are
accelerated in the same way by the Fermi processes. Saying that protons
are accelerated more efficiently than electrons is not true. The only
trouble in ordinary plasmas is that protons are more numerous above the
resonance threshold and the electrons must be efficiently injected
above the threshold to participate to the Fermi processes.

In compact objects, "exotic" plasmas can be created with a copious
relativistic electron (positron) component. The "cauldron" of the black
hole environment (Henri and Pelletier 1991, Marcowith, Pelletier, Henri,
1997) could likely be dominated by the pair plasma. When the most massive
component is due to relativistic electrons (positrons), they are more
numerous above the resonant threshold. Under these interesting
conditions, the power of the acceleration process goes almost entirely in
the radiative particles, which is the best regime to have the most
efficient conversion of energy into radiation.

These exotic plasmas (either relativistic electron dominated with
non relativistic protons or pair dominated) have interesting dynamics.
First, they can be propelled at relativistic velocities by the Compton rocket
effect provided that they are maintained hot in the cauldron (Henri and
Pelletier 1991). Second, the investigation of the nonlinear regime of Alfven
disturbancies (Pelletier, Henri, Marcowith 1997)
shows that acceleration works efficiently only when the magnetic pressure
is larger than the plasma pressure. Overpressured plasmas  (not
confined) suffer radiative cooling and thus come back to rough
equipartition. This regulation process stops the pair creation
catastrophe due to second order Fermi acceleration.
The high energy cut-off is determined by the balance between
the fastest radiation loss rate and the Fermi acceleration rate.
Estimates are given further on.

Note that a plasma dominated by highly relativistic protons is much more
difficult to confine. Proton acceleration is limited by the largest
wavelength of the magnetic disturbancies beyond which scattering no more
works. This is of course in the hot spots of FR2 jets that can be found
the best acceleration sites for protons, and this could be the origin of
the high energy cosmic rays.

\subsection{Time scales}

For a black hole of $10^8$ solar masses, the gravitational radius $r_G$ is of
order of one astronomical unit and a source of radius $100 r_G$ has a
light travel time of $30$ hours. This leads to intraday variability
because the observed time is reduced by the jet Doppler factor:
\begin{equation}
        \tau_{obs} = \frac{\tau_{int}}{{\cal D}} \ .
        \label{TO}
\end{equation}
At a distance of $100 r_G$ from the black hole, a magnetic field of order
$100 Gauss$ can be expected and the photon energy density is comparable
to the magnetic energy density. So the radiation loss time of the
relativistic electrons having a Lorentz factor of $10^3$ is about $100 s$.
A second order Fermi process can easily prevent their energy loss, for
the required pitch angle frequency is only $10^{-8} \omega_s$!.. The TeV
photons would require relativistic electrons with Lorentz factors of
order $10^6$; this is achieved by Fermi process if the pitch angle
frequency is $10^{-3}\omega_s$.

\section{Discussion}

An interesting debate raised to know whether the underlying physics that
explain the gamma emission of blazars is of hadronic or electrodynamic
origin. This focuses on the two issues of
acceleration and variability.

Of course the main argument in favor of the electrodynamic model is that
the electrons allow a much faster variability than protons. The small
size of the high energy sources revealed by their variability would imply
a strong magnetic field to have proton Larmor radii smaller than the size.
It is often unduly said that the electrons are not efficiently
accelerated by Fermi processes, that shocks accelerate more
efficiently than the second order Fermi process, and also that they accelerate
mostly protons. It has been shown that these prejudices are not plainly
true.

The analysis of the second order Fermi acceleration of the relativistic
electrons does not reveal any serious difficulty to explain the gamma
emission of blazars, even to explain the few TeV radiation of the BL-Lacs
(Mrk 421, Mrk 501 etc.). The Klein-Nishina limit seems to be the major
limitation of the inverse Compton emission on accretion disk UV-photons,
the cut-off should be at higher energy for the Synchrotron Self-Compton
emission. Thus the emission of BL-Lacs beyond TeV energy is likely
the SSC-radiation.

In the case of quasars, the inverse Compton process can also be
accompanied by the pair creation process. This seems in fact unavoidable
within $100 r_G$, and it could explain nicely the spectrum break around
few MeV. Indeed only the pair model (Henri, Marcowith, Pelletier 1995) was
able so far to explain a gamma-spectrum index which is twice the X-spectrum
index as observed.

The future multiwavelength compaigns will be crucial to discriminate
between the various models by observing the flares and their growth sequence
in each spectral bands. The delays between each band are predicted
differently by the models.

\section*{Acknowledgments}

G.P. acknowledges the support of his work by the Institut
Universitaire de France.

\end{document}